\documentstyle[floats,aps,prb]{revtex}
\begin{document}
\draft

\twocolumn[\hsize\textwidth\columnwidth\hsize\csname@twocolumnfalse\endcsname

\tighten
\title{Density Functional Theory Calculations of Hopping Rates
of Surface Diffusion}

\author{C. Ratsch\cite{Christian} and M. Scheffler}
\address{Fritz-Haber-Institut der Max-Planck-Gesellschaft, Faradayweg
4-6,
14195 Berlin, Germany}

\date{\today}

\maketitle

\vspace{-.7cm}

\begin{abstract}

Using density-functional theory we compute the energy barriers and
attempt frequencies for surface diffusion of Ag on Ag (111) with 
different lattice constants, and 
on an Ag adsorbate monolayer on Pt (111). We find that the attempt
frequency is of the order of 1 THz for all the systems studied. This is
in contrast to the so-called compensation effect, and to recent 
experimental studies. Our analysis suggests that the applicability of 
simple (commonly used) scaling laws for the determination of diffusion 
and growth parameters is often not valid.
\end{abstract}

\vskip2pc]
 
The determination of the microscopic parameters that govern the 
evolution and morphology of a growing structure during epitaxy 
is the focus of a large number of theoretical and experimental 
studies. However, experimentally deduced values for the diffusion
barriers and attempt frequencies often are not measured directly, 
but result from an analysis of experimental data on the basis of 
some theoretical model. For example, it is {\it qualitatively}
well established that the density of islands $N^{\rm is}$
in growth experiments follows the scaling law~\cite{sto81}
\begin{equation}
N^{\rm is} = C_0 \left( {D \over F} \right)^{-1/3}\quad .
\label{scaling}
\end{equation}
Here $F$ is the deposition flux, $D$ is the diffusion constant, 
and  $C_0$ is a dimensionless quantity that is determined by
geometrical factors and parameters that describe the capture 
efficiency of an island. Recently, Eq.~(\ref{scaling}) has been
applied for a {\it quantitative} determination of diffusion 
constants $D$ for different systems from experimentally measured
island densities~\cite{mo91,str93,bru95}.

Relation (\ref{scaling}) is derived under several assumptions, 
particularly that adatoms are the only species that can diffuse
over the surface, i.e., dimers and larger clusters are immobile, 
and that there is only one mechanism for atomic diffusion, 
typically assumed to be the hop to a nearest-neighbor lattice site.
Then the diffusivity $D$ is related to the hopping rate 
of single adatoms by $D=D_0\,\exp(E_{\rm d}/kT)$,
where $E_{\rm d}$ is the energy barrier for surface diffusion
and $D_0 = \Gamma_0 l^2 / (2\alpha) $ is the pre-exponential factor.
Here  $\Gamma_0$ is
called the attempt frequency, $l$ is the jump length of the diffusing
adatom, and $ \alpha $ reflects the dimensionality
and symmetry of the motion (for a square lattice $\alpha = 2$ ).
It is commonly assumed that $\Gamma_0 \sim 10^{12} - 10^{13}$ 
s$^{-1}$, which is a typical surface phonon frequency. 
We also note that Eq.~(\ref{scaling}) is based on the assumption that
the so-called critical nucleus is $i^*=1$, which means that dimers and 
small clusters  are stable against break-up. This is the case for 
typical growth  experiments as long as the temperature is low enough 
or the deposition rate is high enough \cite{rat94}.

\begin{figure}[tb]
\unitlength1cm
  \begin{picture}(8.0,6.0)
     \includegraphics{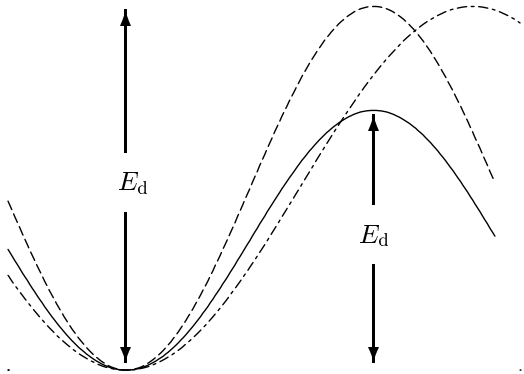}
  \end{picture}
\caption
{Schematic illustration of the compensation effect and the
anti-compensation effect.
When the barrier increases (dashed line compared to solid
line), the potential becomes stiffer, and the vibrational frequencies
increase.
However, when an increase of the barrier is due to an increase in the
lattice
constant, the curvature of the potential might very well
decrease (dashed-dotted line), and as a
result the vibrational frequencies decrease.
In this simplified picture it is assumed that all the vibrational
frequencies
at the saddle point increase (decrease) similarly to the ones at the
lattice
site, and thus all the increases (decreases) of the frequencies
in Eq.~(1) cancel,
except one that corresponds to a vibration in the lattice site.}
\label{anti_compensation}
\end{figure}

In a recent scanning tunneling microscopy (STM) study Brune {\it et 
al.}~\cite{bru95} found that for growth of Ag on a Pt (111) substrate 
the size and density of Ag islands changes significantly when one 
monolayer (ML) of Ag (that grows pseudomorphically on top of the 
Pt substrate) has been deposited.
The density of islands 
in both cases is different again from the density of Ag islands on top 
of clean Ag (111). The authors of Ref.~\cite{bru95}  analyzed their 
experimental data with Eq.~(\ref{scaling}) with the assumption that 
the mechanism for surface diffusion is hopping of a single adatom to a 
nearest neighbor site. They obtained the energy barriers and
attempt frequencies quoted in Table \ref{main_table}, 
\begin{table}[b]
\caption{
Results for  $E_{\rm d}$ and $\Gamma_0$.
All theoretical values are
with $3N=15$, except for Ag on Ag/Pt (111), where $3N=27$.
We considered an Ag(111) substrate with tensile (t), compressive (c),
and without strain. The substrate Ag(111) (c) was compressed to
the lattice constant of Pt.
}
\begin{tabular}{l|cc|ccc}
substrate & \multicolumn{2}{c}{experiment~\cite{bru95}}
&\multicolumn{3}{c}{theory (this work)}\\
&$E_{\rm d}$ (meV) & $\Gamma_0$ (THz)
&$E_{\rm d}$ (meV) & $\Gamma_0$ (THz) \\
\hline
Ag (111) (t)     &  --    & --      &  106  &  0.25 \\
Ag (111)           &  100   &  0.2    &  82   &  0.82 \\
Ag (111) (c)     &   --  &   --     &  60   &  1.3 \\
Ag/Pt (111)        &  60    &  0.001  &  60   &  5.9 \\
Pt (111)           &  160   & 10      &  150  &   --
\end{tabular}
\label{main_table}
\end{table}
which shows that the attempt frequency found for Ag on 1 ML Ag on
Pt(111)
is much smaller than that for Ag on Ag (111). A lower attempt frequency 
for a system with a lower diffusion barrier is not an unexpected result, 
and is known in the literature as the compensation effect~\cite{mey37}. 
It can be understood with simple physical arguments that are illustrated 
in Fig.~\ref{anti_compensation}.
Consider the solid and the dashed line
that
are the potential energy curves for a system with a low and with a high
barrier. It is evident that the curvature of the potential with the
higher barrier is larger at the bottom of the potential well. An atom in
a well with a higher curvature vibrates faster, and 
thus the attempt frequency is larger.
 
In a molecular dynamics study~\cite{boi95} that employed empirical
potentials and the embedded-atom theory
it was found that indeed for homoepitaxial diffusion on different metal 
surfaces there is a systematic compensation 
effect, i.e., systems that have a larger energy barrier for surface
diffusion have a larger attempt frequency.
Values for $\Gamma_0$ are obtained that vary from $3 \times 10^{12}$
s$^{-1}$
to $5 \times 10^{13}$ s$^{-1}$, while for the same systems
the energy barriers for diffusion increase from
$0.05$ eV to $0.7$ eV. The magnitude of the effect for $\Gamma_0$
is in the range of typical surface phonon frequencies, and it does not 
compensate completely (for typical growth temperatures) the increased 
diffusion barrier. In particular, we note that the attempt frequency is
always much larger than 
the reported value of $\Gamma_0 = 10^9$ s$^{-1}$ for Ag on 1ML Ag on
Pt(111), which is about three {\em orders of magnitude} smaller than
what one would expect.
Despite this puzzle, we note that previous theoretical studies seem
to suggest that the system 1 ML Ag on Pt (111) behaves rather normal
and that, with respect to surface diffusion, it behaves
essentially like a laterally compressed Ag (111)
substrate~\cite{bru95,rat97}.

The diffusion constant is determined by the diffusion barrier and 
pre-exponential factor, but for the latter no {\it ab initio}
calculations have been reported so far. This applies 
in general, not just to the system of concern in this Letter.
In view of the above mentioned puzzle, and similar enigma that 
seem to exist for other systems, we performed density functional theory 
calculations of the attempt frequencies for Ag diffusion on Ag (111) 
with different lattice constants, and
on 1 ML adsorbed on Pt (111).
We find that in all cases the attempt frequency is of the order of 
1 THz, so that claims of a large variation of $\Gamma_0$ 
are not supported. We then discuss various mechanisms
which become operative in systems with low energy barriers.

We employ density-functional theory with the local-density 
approximation \cite{fhi96md}. More details are given in
Ref.~\cite{rat97}.
Unless noted otherwise, we use a supercell with 4 layers, 
a ($2 \times 2$) surface cell, and the adatom 
placed on only one side of the slab. Tests have shown
that such a cell is large enough~\cite{rat97}.

The energetically most favored adsorption sites 
are the three-fold coordinated fcc and hcp sites, and an adatom that 
diffuses from one site to another will pass over the bridge site
as the transition state \cite{true_saddle}. 
In principle, the effective diffusion constant on a surface with 
several adsorption sites within one unit cell needs to be calculated as a
combination of the different rates for the motion between the different 
sites. However, for all the systems studied here the fcc site is always 
slightly favored over the hcp site (by 5 -- 10 meV),
but the barrier to hop from an hcp site 
to an fcc site is almost identical to the one to hop from an fcc site to
an hcp site.
This provides justification that only the fcc site is considered as the 
(stable) adsorption site.
Transition-state theory (TST) within the harmonic approximation is used,
and the
attempt frequency is calculated from~\cite{vin57}
\begin{equation}
\Gamma_0 = \frac{\prod_{j = 1}^{3N} \, \nu_j}{\prod_{j = 1}^{3N-1}
\,\nu_j^{*}} \quad .
\label{vinyard}
\end{equation}
The $\nu_j$ ($\nu_j^{*}$) are the normal-mode frequencies of the system
with the adatom at the fcc site (bridge site), and $3N$ is
the number of degrees of freedom of the system.

We have calculated these normal mode frequencies by 
first evaluating and then diagonalizing the 
force-constant matrix. The elements of this matrix are obtained by
displacing 
each individual atom in the cell and calculating the forces that act on
all the atoms. An important issue is how large $3N$ should be
for constructing the force-constant matrix. 
The results obtained for different $3N$ are summarized in 
Table~\ref{convergence} 
\begin{table}[tb]
\caption{
The calculated attempt frequency $\Gamma_0$ for Ag on Ag(111) for
different numbers of degree of freedom considered.
}
\begin{tabular}{lccc}
$3N$ & 3 & 15 & 99 \\
\hline
$\Gamma_0$ (THz) & 1.55 & 0.82 & 0.71 \\
\end{tabular}
\label{convergence}
\end{table}
for Ag on Ag (111). If we change $3N$
from 3 to 15, i.e., take into account the adatom and all atoms in the
first layer, the attempt frequency changes from 1.55 THz to 0.82 THz.
We also developed a scheme to include a larger number of 
degrees of freedom, and given in Table~\ref{convergence}
is a value for $3N = 99$ which corresponds to a ($4 \times 4$) surface 
cell with vibrations of the adatoms and two layers of the substrate.
The $(99 \times 99)$ dynamical matrix was constructed
as follows: We use the same force constants as in the $3N=15$
situation in the corresponding subspace. For the other
force constants we use bulk values (for atoms in the second and deeper
layer), and clean-surface values (for farther surface atoms).
Tests with a ($3 \times 3$) cell revealed that force constants of
surface atoms due to the displacement of atoms that are further
away than nearest neighbors
are negligible. Similarly, force constants of atoms in the second
layer (and below) that result from the displacement of atoms that are
further away than its nearest neighbors can be neglected. 
We estimate that the influence of these negligible force constants 
on the attempt frequency is less than 
a factor two, and therefore set them to zero.
The attempt frequency for $3N = 99$ decreases very little compared
to the $3N =15$ treatment, namely to 0.71 THz, and thus
our result obtained when only the adatom is allowed to vibrate
is correct within a factor of two.

As argued earlier \cite{bru95,rat97}, the main difference between
an Ag atom on a clean Ag (111) surface and on 1 ML Ag on Pt (111) 
is due to strain in the Ag adlayer. We therefore studied
the attempt frequencies for Ag on Ag (111) as a function of strain. The
results are given in Table~\ref{main_table}.
An increasing lattice 
constant implies that the diffusion barrier increases~\cite{rat97}, so, 
according to the compensation effect, the attempt frequency should
increase.
Surprisingly, the opposite is the case: When the barrier increases from
$60$ meV to $106$ meV, the attempt frequency decreases from $1.3$ THz
to $0.25$ THz. We 
understand this ``wrong  trend'' as follows:
When the diffusion barrier increases because the lattice constant 
increases, the curvature of the potential decreases,
so that the  attempt frequency decreases. This anti-compensating
effect is sketched schematically in 
Fig.~\ref{anti_compensation} by the dashed-dotted line. 

Our result for Ag on unstrained Ag (111) differs by a factor 
of $\sim 4$
from that obtained in the analysis of the experimental data.
This difference is partially due to the fact that the analysis of the
experimental data ignored that an atom that hops from an fcc
site to a neighbor fcc site first visits an hcp site. When the 
difference in adsorption energies between the fcc and hcp sites
is very small, the adatom needs to make (on average) three times as many 
hopping attempts to get to one of the six adjacent fcc sites than it 
needs to hop to one of the three adjacent hcp sites. Thus, the attempt 
frequencies given in Ref.~\onlinecite{bru95}, which are also noted in 
Table \ref{main_table}, should be multiplied by a factor of three.

For the diffusion of an Ag adatom on 1 ML Ag adlayer on Pt (111)
the experimentally deduced attempt frequency is 
more than two {\em orders of magnitude}
smaller than that for the Ag (111) surface, but such a reduction 
is not found in our calculations (compare Table \ref{main_table}).
When only the adatom and the Ag layer are included
we obtain $\Gamma_0 \simeq 3.8$ THz ($3N = 15$), and upon inclusion of
the top Pt layer we obtain $\Gamma_0 \simeq 5.9$ THz ($3N = 27$).
As mentioned earlier, with respect to diffusion energy barriers,
the Ag monolayer on Pt (111) behaves very much like a compressed
Ag (111) surface. We now find that this also holds for the 
attempt frequency.
The disagreement with the experimentally deduced value,
that is three order of magnitude smaller, is  significant, and we
consider this disagreement a most important result. It demonstrates that
the understanding of surface diffusion and growth is far from being
complete, although often this is not appreciated. 
In this regard we note that Eq.~(\ref{scaling}) is derived
under several assumptions, and that often some of them are not met.
We will now discuss six different problems which may 
have troubled the experimental analysis and/or our theoretical results.

$i):$ It is a concern whether the harmonic approximation we used 
is valid for all the systems under consideration. From the calculation
of the forces we find that for the systems studied here
the harmonic approximation is indeed a rather good approximation
for displacements up to $0.2$ \AA.

$ii):$ It is known that TST is only working when the following
conditions are met: The adatom spends enough time 
at the adsorption site so that it can
equilibrate there. This aspect is connected to the phononic friction,
i.e., the strength of the coupling of the adatom vibration to the 
excitations of the substrate.
Additionally, once an atom has crossed the transition site 
there should be no recrossing before the atom has 
been equilibrated again. While it is clear that the 
friction should not be too small or too large,
a quantitative criterion when TST is valid is not known~\cite{thumb}.
We expect that equilibrium might not be reached when the temperature is
too high, but the experiments of Ref.~\cite{bru95} where carried out at 
temperatures that are smaller than half of the Debye temperature.

$iii):$ 
Diffusion of small clusters on a metal surface can be 
significant \cite{wen94}, but Eq.~(\ref{scaling}) is only valid when the
only mobile species are adatoms. In fact, it has been shown
~\cite{vil92} 
that the scaling exponent in Eq.~(\ref{scaling}) changes when small
clusters are allowed to move and that the island density decreases 
dramatically~\cite{kui96}. This effect cannot explain the 
discrepancy we find for the  diffusion on Ag/Pt (111):
If cluster diffusion were important on Ag/Pt(111)
and not (or less) on clean
Ag (111), the pre-exponential factor for diffusion on
Ag/Pt(111) should be bigger, not
smaller, as observed in the experiments.
Still, we do not
rule out that cluster diffusion plays a role, because 
it is possible and likely that the diffusivity
of cluster diffusion is (very) different to that of single-atom
diffusion, and we also note that the exponent in Eq.~(\ref{scaling})
may change.
We are not aware of the direction and size of this 
effect, but note that cluster diffusion may be particularly 
important for strained systems and thus Ag diffusion on Ag/Pt(111)
is indeed a good candidate.

$iv):$ It has been shown that on metal surfaces the diffusion length may
be a multiple of the lateral lattice constant~\cite{sen95,lin97}. This 
becomes more relevant when the corrugation of the potential energy
surface 
is less pronounced and the diffusion barrier becomes small~\cite{zero},
as it is the case for Ag on Ag/Pt(111) and Ag on a compressed Ag (111)
substrate. The adatom becomes more delocalized and we speculate that
rather than hopping over the surface from one
lattice site to an adjacent one, the
adatom is ``flying'' over the surface over several lattice sites. 
For this process TST is strictly no longer applicable.
Moreover, it is conceivable that Eq.~(\ref{scaling}), particularly 
its exponent, changes when several diffusion mechanism are operative.
We are not aware of the size or direction of such an effect,
but under the speculative assumption that the exponent is reduced by 
20\%, the discrepancy between the results of Ref.~\cite{bru95} and 
our calculations could be explained. The analysis of the experimentally 
measured island density would then yield an activation barrier that 
it 20\% larger, which is supported by experimental~\cite{lin97}
and theoretical~\cite{jac97} results for the double-jump of a Pt adatom
on Pt(110). This would also explain the good agreement between our 
calculated and the experimentally deduced diffusion barriers.
Additionally, an exponent that it 20\% smaller yields an attempt
frequency of 1 THz, in agreement with our calculated value.

$v):$ We also note the possibility of diffusion by
atomic exchange over long distances. The Ag adlayer is compressed and
we speculate that atomic exchange could work such that an atom of the Ag 
adlayer pops up, becoming an adsorbate atom. Then the vacancy 
moves towards another adatom (e.g. a long row of Ag adatoms in the
Ag adlayer may move).
Similar effects have been observed in embedded atom simulations
\cite{papanik}.

$vi):$ Another problem is knowledge of the dimensionless quantity $C_0$
in eq. (\ref{scaling}). An uncertainty in $C_0$ of one order of 
magnitude translates into an uncertainty for $\Gamma_0$ of three orders
of magnitude when $i^*=1$.
The value of $C_0$ is essentially determined by the surface geometry and
capture efficiency of islands, and 
with the self consistent approach of Bales and Chrzan~\cite{bal94} one 
obtains $C_0 \simeq 0.25$, which has been used in 
Ref.~\cite{bru95}. 
A recent STM study for Pt on Pt(111) used
an alternative method to determine $E_{\rm d}$ and
$\Gamma_0$~\cite{bot96}.
It is found that $\Gamma_0=5$ THz (an expected result), and from the
data in~\cite{bot96} we deduce $C_0 \simeq 0.28$.
Earlier work~\cite{sto71} suggested that $C_0$
should be of the order of unity.
It is known that often there is
{\it funneling} of adatoms toward step edges. This attraction 
will become more significant for systems with low diffusion barriers [as
it is the case for Ag on 1 ML Ag on Pt (111)], and when surface states
are important and affected by adatoms and/or steps.

We conclude that attempt frequencies of single-atom hopping diffusion
are indeed of the order of 1 THz and reported deviations from this value,
as for example the reduction by three orders of magnitude~\cite{bru95},
are only apparent and actuated by
the low energy barrier of diffusion. This implies that adatoms
should be described by a more delocalized nuclear wavefunction.
Long-range adatom-adatom and atom-step interactions become noticeable,
and atomic diffusion proceeds more by ``flying'' over several
lattice constants rather than by hopping to nearest neighbor
sites. For strained systems also cluster diffusion may be
enhanced. The importance of these effects is difficult to
assess by theory, but we trust that more experiments
will help to clarify the situation.

We acknowledge many helpful discussions with 
H. Brune, K. Fichthorn, K. Kern, A. Kley, 
P. Ruggerone, and A.P. Seitsonen.


\begin{references}
\bibitem{true_saddle}
Because the bridge site separates an fcc and an hcp adsorption site, the 
true saddle point is slightly displaced from the bridge site, but 
we find that in the system studied here this asymmetry is irrelevant.
\bibitem{thumb}
As a rule of thumb TST is justified when
$E_{\rm d} >> kT$.
\bibitem{zero}
TST might not be reliable and long jumps become important
when the energy of zero-point vibrations 
is comparable to the energy barrier. For the systems studied here, we 
estimate this energy to be about 15 meV.
\end{references}
\end{document}